\documentclass{nature}
\usepackage{indentfirst}
\usepackage{graphicx}
\usepackage{color}
\usepackage{amssymb}
\usepackage{amsmath}

\title{Yu-Shiba-Rusinov bound states of exciton condensate}

\author{SeongJin Kwon$^{1,2}$, Kyung-Hwan Jin$^{1,7}$, Jong Eun Han$^{3}$, Siwon Lee$^{1,2}$, ChoongJae Won$^{4,5}$, Sang-Wook Cheong$^{4,5,6}$, Han Woong Yeom$^{1,2\dagger}$}

\begin{document}

\maketitle

\begin{affiliations}
 \item Center for Artificial Low Dimensional Electronic Systems, Institute for Basic Science, Pohang 37673, Korea.
 \item Department of Physics, Pohang University of Science and Technology, Pohang 37673, Korea.
 \item Department of Physics, University at Buffalo, State University of New York, Buffalo, New York 14260, USA.
 \item Laboratory for Pohang Emergent Materials, POSTECH, Pohang 37673, Korea
 \item MPPC-CPM, Max Planck POSTECH/Korea Research Initiative, Pohang 37673, Korea
 \item Keck Center for Quantum Magnetism and Department of Physics and Astronomy, Piscataway, New Jersey 08854, United States
 \item Department of Physics and Research Institute of Materials and Energy Sciences, Jeonbuk National University, Jeonju, 54896 Republic of Korea
 
 $^{\dagger}$yeom@postech.ac.kr
 
\end{affiliations}

\begin{abstract}
Quantum condensed states in solids often reveal their fundamental nature via interactions with impurities, as epitomized by Yu-Shiba-Rusinov (YSR) bound states at magnetic impurities in superconductors. Although analogous YSR bound states were predicted within quantum condensates of excitons several decades ago, their existence has been elusive. Here, we directly visualize in-gap electronic states bound to impurities inside an exciton condensate phase of a van der Waals crystal Ta$_2$Pd$_3$Te$_5$, utilizing scanning tunneling microscopy and spectroscopy. 
We find that the energies of in-gap states are strongly correlated with the excitonic band gap, which is systematically tuned by local strain and carrier injection. Our theoretical analyses reveal that these in-gap states are induced by charge dipoles associated with Ta vacancies through a charge-exciton version of the YSR mechanism. 
Our findings establish both the YSR physics in exciton condensates and a novel microscopic tool to probe and control quantum properties in exciton condensates persisting up to room temperature.
\end{abstract}

\textbf{Introduction} - Quantum condensates are found in solids and cold atoms as quantum mechanically coherent states of quasipaticles or atoms exhibiting macroscopic quantum phenomena such as superfluidicity and superconductivity. 
Their many-body quantum nature is often exhibited through interactions with specific impurities.
A prominent example is the Yu-Shiba-Rusinov (YSR) resonance in superconductors, which arises from the many-body interaction between a Cooper pair condensate and a magnetic impurity \cite {YSR1, YSR2, YSR3}.
This interaction locally disrupts Cooper pairing via a time-reversal-symmetry-breaking by a local spin, resulting in a resonance state at the impurity site within the superconducting gap \cite{YSRexample1, YSRexample2, YSRexample3, YSRexample4}.
The YSR states are sensitive probes to many-body interactions in superconductors and have provided profound insights into the pairing mechanism and the interplay between magnetism and superconductivity \cite{YSRexample3, YSRexample4, YSRexample5, YSRexample6, YSRexample7, YSRexample8, YSRexample9, YSRexample10, YSRexample11}.
Their presence is proximal to Majorana zero modes in topological superconductors, making them extremely important in unambiguous identification of Majorana zero modes \cite{Majorana1, Majorana2, Majorana3}.
Moreover, magnetic impurities can be widely manipulated, leading to the engineering of quantum phase transitions and to designed quantum states \cite{YSRmanipulation1, YSRmanipulation2, YSRmanipulation3, YSRmanipulation4}.

Although not detected experimentally yet, a YSR resonance can exist in other types of condensates of bosonic quasiparticles. 
For example, the YSR theory predicted long time ago that a charged impurity could locally break electron-hole pairing, forming impurity bound states within a quantum condensate of excitons, so called exciton insulator \cite{exciton1, exciton2, exciton3, exciton4}.
Despite recent establishments and promising applications of various forms of exciton condensates including electron-hole bilayers in semiconductors \cite{bilayer1, bilayer2, bilayer3, bilayer4, bilayer5, bilayer6}, graphene \cite{graphene1, graphene2}, transition metal dichalcogenides (TMDCs) \cite{TiSe21, TiSe22, Ta2NiSe51, Ta2NiSe52}, as well as in simple semimetallic TMDC monolayers \cite{monolayer1, monolayer2} and three-dimensional semimetallic crystals \cite{carbonnanotube, MoS2}, the YSR impurity-bound state in an exciton condensate has remained to be discovered.

Here, we report the observation of characteristic in-gap electronic states locally around defects in Ta$_2$Pd$_3$Te$_5$, which was reported to become an exciton insulator at and below room temperature \cite{TPT1, TPT2, TPT3, TPT4}.
Pairs of in-gap states around specific intrinsic defects, bearing striking similarities to superconducting YSR states, are observed.
Local control of the strain and the carrier density established a solid relationship between the defect-bound states and the excitonic order.
Computational results suggest that the defects are Ta vacancies with a local charge dipole, which produce a double resonance within the excitonic gap.

\textbf{Atomic structures and excitonic phase transition} - Ta$_2$Pd$_3$Te$_5$ is an orthorhombic van der Waals crystal with lattice constants of $a=13.9\,\mathrm{\text{\AA}}$, $b=3.76\,\mathrm{\text{\AA}}$, $c=18.72\,\mathrm{\text{\AA}}$ \cite{TPTstructure}.
Each unit cell contains two Ta$_2$Pd$_3$Te$_5$ layers, which are composed of alternating one-dimensional chains of Ta and Pd (Fig. 1a.) sandwiched by Te layers.
Te atoms form an alternating pattern of dimer and trimer chains along the c-axis, as clearly resolved in STM topographies (Fig. 1b) and closely simulated in DFT calculations (Fig. 1c and Supplementary Fig. S1 for the electronic band structure).
The transition from a zero-gap semimetal to an insulator with a gap of 100 meV was observed below 365 K by transport measurements\cite{TPT1, TPT2} and angle-resolved photoemission spectroscopy (ARPES) \cite{TPT1, TPT2, TPT4, TPT9}.
The structural distortion accompanying the exciton condensation was found to be too small to be detected in x-ray diffraction and electron microscopy \cite{TPT1, TPT2, TPT6, TPT7}. 
The excitonic nature of the transition was recently confirmed by identifying the direct exciton photoemission signal in ARPES \cite{TPT4}.

ARPES and DFT calculations consistently identified one conduction (CB1) and two valence bands (VB1, VB2) near the Fermi energy (Fig. 1d and Supplementary Fig. S1) \cite{TPT3}. 
The CB1 and VB1 bands form a Dirac crossing to make the system metallic in the normal phase. 
The electrons and holes in these two bands, respectively, are expected to form excitons to open a many-body energy gap below 370 K \cite{TPT3}. 
The band gap is fully saturated with a size of about 100 meV as shown in the ARPES spectra at 80 K and can also be seen in scanning tunneling spectroscopy (STS) spectra at 4.4 K.
Note that the gap and the local density of states (LDOS) across the Fermi level are asymmetric mainly due to the presence of an extra valence band VB2.  

\textbf{Defects and in-gap states }- Lowering the bias in the STM measurements to near the excitonic gap region ($\leq$100 mV), we observe atomic scale defects, appearing as either strong protrusions or much weaker spots in the topography (Fig. 1f and Supplementary Fig. S2).
The LDOS maps inside the gap, both in filled and empty states, reveal strong in-gap electronic states at positions of the protrusion-type defects (Fig. 1g). 
In sharp contrast, defects of the other type do not show any in-gap state (See Supplementary Fig. S3).
One can easily notice that the defects are indeed in atomic size in the enlarged topographies in Fig. 2 but the in-gap states have much wider extents, up to 6 nm in the LDOS maps. 
This immediately suggests that the wavefunction of the in-gap state arises from some many-body interaction, as opposed to local atomic orbitals on defect sites. 
The detailed dI/dV spectra (Fig. 2) confirm two electronic states split from the edges of valence and conduction bands, respectively, around the defects. 
The detailed STM images reveal five different defect sites (called $\alpha$,$\alpha'$,$\beta$,$\beta'$, and $\gamma$) for the defects with in-gap states (Supplementary Fig. S4).
While they have marginally different energies, their in-gap states share the same essential characteristics with a pair of sharp peaks split from the valence and conduction band edges, in a similar manner as the YSR states of magnetic impurities in superconductors (Supplementary Fig. S5).
The existence of impurity-induced in-gap states in the filled state was previously suggested by the transport\cite{TPT6} and ARPES measurements\cite{TPT2, TPT6}. 

The defects showing in-gap states feature atomic-scale depressions in one Te chain and a strong protrusion in the neighboring chain. 
These topographic data were input for extensive modeling using DFT calculations based on structural defects such as Te, Pd, and Ta vacancies (see Supplementary Fig. S6 and S7).
The calculations reveal that Ta and Pd vacancies on the topmost layer explain the defects with and without the in-gap states, respectively.
Figures 3a and 3b compare the experimental and simulated STM images of one Ta vacancy structure. 
Since there are four Ta chains within a single unit cell (Fig. 1a), four different Ta vacancy sites are possible to explain $\alpha$, $\alpha'$, $\beta$, and $\beta'$ defects.
$\alpha$ and $\alpha'$ ($\beta$ and $\beta'$) are located in mirror-symmetric sites as reflected in their symmetric STM images. 
Due to the lack of a strong topographic contrast and small absolute spectral intensity of in-gap states, $\gamma$ defects are thought to be the other defects in the layer underneath. 
While the STM images of all four defects are reasonably well reproduced by the DFT simulations (Supplementary Fig. S8), the DFT calculations do not predict the pairs of in-gap LDOS for any of these defects.
Moreover, all the electronic states localized on defects have spatial extents of less than 1 nm, in stark contrast to the in-gap states observed.

\textbf{Tuning excitonic order }- We thus hypothesized that these in-gap states are due to the interaction with the exciton quantum condensate and examined how they evolve upon the change of the excitonic order parameter, namely, the excitonic band gap. 
We note that the excitonic band gap is sensitive to strain as observed in the transport measurement under pressure \cite{TPT8}.  
As shown in Fig. 3a and c, we found a corrugated stripe region with a width of 8 nm, which exhibits a strongly strained lattice locally (strain variation of ${\epsilon}\,{\approx}\,$2.4\%, Supplementary Fig. S9).
In this region, the excitonic gap collapses into a metallic pseudogap of an order of 30 meV (the grey spectra in Fig. 3d).
The pseudogap indicates strong excitonic fluctuation, the preformed excitons, which is expected for a Bose-Einstein condensation transition \cite{preformedexciton1, preformedexciton2, preformendexciton3}.
A few nontrivial defects were found within the stripe as shown in Fig. 3b.
They, however, do not show any signatures of an in-gap level inside the pseudogap (Fig. 3d). 
This clearly indicates that the in-gap states originate from the interaction of the defects with the excitonic order.

Inspired by recent STM studies on another exciton insulator candidate Ta$_2$NiSe$_5$ \cite{Gapcontrol1, Gapcontrol2, Gapcontrol3}, we systematically tuned the excitonic gap by increasing the transient carrier doping induced by the tunneling current. 
Figure 3e shows the successful demonstration of the excitonic band gap control by increasing (decreasing) tunneling current (tip height) on a defect-free area. 
The apparent gap size changes from about 140 to 60 meV until the tip virtually touches the surface
(Supplementary Figs. S10 and S11).
Then, the band gap abruptly collapses into a metallic pseudogapped state, which is consistent with that observed in the strained region. 
The gap itself and its variation are asymmetric with respect to the Fermi energy. 
The rather robust band edge in the filled state must be due to the presence of an extra spectator valence band VB2, which defines the valence band edge in the exciton insulator phase. 
We performed the same gap tuning on a few non-trivial defects. 
In the course of controlled gap-decrease, the energy of the in-gap state (especially E$^+$) of an $\alpha$ defect is systematically reduced until it disappears (E$^-$) or decays (E$^+$) in the pseudogap metallic state (Figs. 3e and 3f). 
Similar behaviors are observed for all non-trivial defects with in-gap states as summarized in Fig. 3g (See Supplementary Fig. S12).
This result is consistent with the above strained region data and solidly confirms the excitonic origin of the in-gap states. 

\textbf{YSR theory} - The excitonic YSR theory predicts that a charged impurity induces in-gap states inside the excitonic gap since the Coulomb interaction for the exciton formation is readily perturbed by charges, in direct analogy to magnetic impurities for the spin-singlet Cooper pairs.
Our DFT calculations tell that the non-trivial Ta-vacancy defects have local charges (Fig. 4d) in sharp contrast to trivial Pd-vacancy defects (Supplementary Fig. S13), suggesting that the present in-gap states are related to charges. 
In more detail, the excess local charges of a Ta defect form a dipole, aligned perpendicular to the chain direction with a dipole moment of $p_z=-0.45\times10^{-12}$ C${\cdot}$m (Fig. 4d, e).
Motivated by the DFT results, we performed a model calculation within the YSR theory for a simple symmetric dipole and a monopole.
A two-dimensional tight-binding Hamiltonian of a gapped system due to excitonic pairing was set up (See Methods section) with a local perturbation term of a charge monopole and dipole. For simplicity, we set up the electron and hole bands as electron-hole symmetric.
Our calculations reveal that the local charge dipole gives rise to two in-gap bound state solutions separated in filled and empty states (Fig. 4f, top), while the single charge results in in-gap states only in filled states (Fig. 4f, bottom).
When the dipole potential is sufficiently weak, two impurity bound states symmetrically exist at the continuum energy level.
However, as the potential strengthens, the resonance states segregate from the continuum toward the Fermi level in qualitative agreement with our observation.
The main difference from theory exists in the experimental spectral asymmetry of the two in-gap states (Figs. 3e and 3g).
We suggest that this discrepancy comes from the asymmetry of electron and hole bands due partly to the presence of an extra valence band.
In general, an excitonic insulator does not necessarily respect the electron-hole symmetry in contrast to a superconductor \cite{asymmetry1, asymmetry2, asymmetry3}, which would readily be reflected in the behaviors of YSR states.

While we relate the charge dipole to the defect-bound states, any pair-breaking local interaction can induce similar bound states. 
One obvious candidate is the local strain as suggested in Fig. 3. 
However, we think such a possibility is low since a non-charged Pd vacancy does not bind any in-gap state. 
Another possibility to explain the defect-bound state is the quantum-spin Hall bound state in a point defect~\cite{TPT10, TPT11}, given the proposal of a topological excitonic insulator in the present crystal. 
This possibility is not likely since we do not see any splitting of the bound state up to a magnetic field of 2.5 Tesla (Supplementary Fig. S14).

\textbf{Conclusions - }We have discovered the local electronic states bound to impurities within the excitonic band gap of an excitonic insulator and interpreted that they originate from the pair breaking interactions of a charge-dipole impurity. The presence of the interaction-driven impurity-bound states in turn manifests the excitonic nature of the present insulating phase. 
A charged impurity in an excitonic insulator would play the role of magnetic impurities in superconductors, providing a versatile probe to investigate exciton quantum matters and to engineer them in a wider variety of artificial and natural exciton condensates given the promising applications of exciton condensates in dissipation-less devices and coherent optoelectronics. Such an opportunity is open even to room temperature in a few exciton condensates including the present system with extremely high condensation temperatures realized.



\clearpage

\begin{methods}
\textbf{Crystal growth}
Single crystals of Ta$_2$Pd$_3$Te$_5$ were prepared via the self-flux synthesis.
A mixture of Ta (powder, 99.999 \%), Pd (powder, 99.9999 \%), and Te (lump, 99.9999 \%) materials in a molar ratio of 2.0 : 4.5 : 7.5 was placed in an alumina crucible and heated to 950 $^{\circ}\mathrm{C}$ over a period of 10 hours and sustained for 2 days.
The crucible was then cooled to 800 $^{\circ}\mathrm{C}$ with a temperature change rate of 0.5 $^{\circ}\mathrm{C}/h$.
The crystalline quality was confirmed via both X-ray diffraction and spectroscopy (Supplementary Fig. S15 and S16).

\textbf{STM measurements}
Experimental measurements on Ta$_2$Pd$_3$Te$_5$ were conducted using a commercial cryogenic Joule-Thompson (JT) STM (SPECS, Germany). 
Ta$_2$Pd$_3$Te$_5$ crystals were cleaved at ultra-high vacuum condition below 5$\times$10$^{-8}$ Torr and immediately transferred into the STM stage to ensure a clean surface.
All STM and STS measurements were performed at 4.4 K using mechanically cut PtIr tip.

\textbf{ARPES experiment}
Ta$_2$Pd$_3$Te$_5$ crystals were cleaved at 80 K in UHV condition below 2$\times$10$^{-10}$ Torr before measurements.
ARPES experiments were performed using the commercial DA-30L electron analyzer (Scienta Omicron, Sweden) and a laser of photon energy 11 eV (Oxide, Japan), with the energy resolution better than 0.6 meV.

\textbf{Computations}
The first-principles calculations were performed within the framework of the density functional theory (DFT) using the Perdew-Burke-Ernzerhof-type generalized gradient approximation for the exchange correlation functional, as implemented in the Vienna ab initio simulation package \cite{DFT1, DFT2}.
In the self-consistent process, all the calculations were carried out using the kinetic energy cutoff of 520 eV on a 15×3×4 Monkhorst-Pack k-point mesh for Ta$_2$Pd$_3$Te$_5$ bulk calculations.
The spin–orbit coupling effect was included in the self-consistent electronic structure calculation. 
The lattice constants were taken from experiments \cite{TPT11} with the atoms in the unit cell fully relaxed with the force cutoff of 0.01 eV $\AA$$^{-1}$.
The electronic self-consistent iteration was converged to 10$^{-5}$ eV precision of the total energy.

With the experimental and DFT parameters as input, we set up two-dimensional $41\times41$ square lattice tight-binding Hamiltonian of a gapped excitonic insulator as:
\begin{equation}
\hat{H} = -t \sum_{\langle ij \rangle} c_i^\dagger c_j - \mu \sum_i c_i^\dagger c_i +t \sum_{\langle ij \rangle} v_i^\dagger v_j + \mu \sum_i v_i^\dagger v_i \\
- \Delta \sum_i (c_i^\dagger v_i + v_i^\dagger c_i) + \sum_{i} V_i (c_i^\dagger c_i + v_i^\dagger v_i) 
\end{equation}

The first 2 terms describe the tight-binding model for the conduction orbitals represented by $c_i$ at site $i$, with the hopping parameter $t$ and the chemical potential $\mu$. 
The valence orbitals $v_i$ are similarly defined with parameters of opposite signs of the conduction band.
The overlap $\Delta$ in the 5th term represents the uniform coupling between the conduction and valence bands mediated by the excitonic condensate.
The model parameters are $t=0.5$ eV, $\mu=-1.5$ eV, and $\Delta=50$ meV.
The lattice constant of the system is set to 0.25 nm.
The last term introduces the electrostatic potential $V_i$ from the dipole impurity.
The dipole potential is modeled by two displaced Gaussian peaks of strength $V_0$ and width $\sigma=2$ nm, located at $(\pm\frac12d,0)$ with $d=1$ nm.
$V_0$ values are shown in Fig. 4f.

\begin{equation}
V_i^{dipole} = {V_0} \left( \exp \left[ - \frac{(x_i - \frac d2)^2 + y_i^2}{2\sigma^2} \right] - \exp \left[ - \frac{(x_i + \frac d2)^2 + y_i^2}{2\sigma^2} \right] \right)
\end{equation}

while the monopole potential is described as
\begin{equation}
V_i^{mono} = {V_0} \left( \exp \left[ - \frac{x_i^2 + y_i^2}{2\sigma^2} \right] \right)
\end{equation}

\end{methods}

\begin{addendum}
\item[Data availability]  The source data used for Figs. 1, 2, 3, and 4 are fully available on request
from the authors and are provided as in Source Data file with this paper.

\end{addendum}

\bibliography{Reference.bib}

\begin{addendum}
 \item This work was supported by the Institute for Basic Science (Grant No. IBS-R014-D1).
 K.-H.J. was supported by Global-Learning \& Academic research institution for Master's·PhD students, and Postdocs (LAMP) Program of the National Research Foundation of Korea (NRF) funded by the Ministry of Education (Grant No. RS-2024-00443714).
 C.W. was supported by NRF research grant (No. RS-2022-NR068223). 
 S-W.C. was supported by the W. M. Keck Foundation grant to the Keck Center for Quantum Magnetism at Rutgers University.
 H.W.Y. appreciates the discussion with Gil Young Cho. 
 
 \item[Author contributions] S.K. performed STM experiments. S.L. performed ARPES experiments. K.-H.J. performed DFT calculations and simulations. J.E.H. performed model calculations. C.J.W. synthesized the crystal under the supervision of S.W.C. H.W.Y. conceived the main ideas and supervised the whole project. S.K., J.E.H., and H.W.Y. wrote the manuscript.
 \item[Competing Interests] The authors declare no competing interests.
 \item[Correspondence] Correspondence and requests for materials should be addressed to H. W. Yeom (email: yeom@postech.ac.kr).

\end{addendum}

\clearpage

\begin{figure*}[htp!]%
\centering
\includegraphics[width=0.9\textwidth]{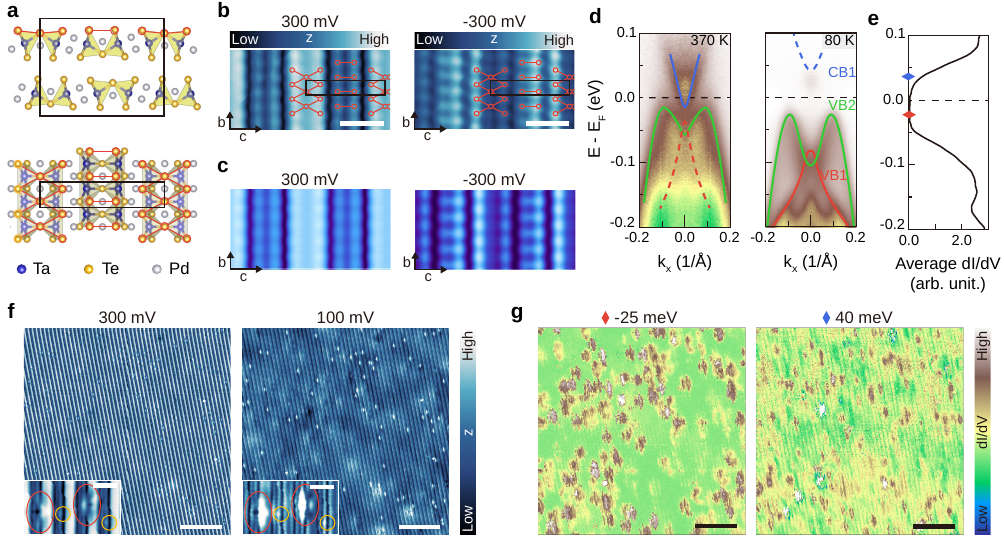}
\caption{\textbf{Atomic structures, excitonic transition, and defects of Ta$_2$Pd$_3$Te$_5$.}
\textbf{a.} Top and side views of the schematic atomic structure. Topmost Te atoms and their bonding are highlighted by red circles and lines, respectively.
\textbf{b.} STM topography images obtained at sample bias of 300 mV and -300 mV. The structure of topmost Te atoms is overlaid with red circles and lines. The lattice unit cell is indicated by a black box.
Scale bars represent 1 nm.
\textbf{c.} STM topographies simulated by DFT calculations in the given bias conditions match well with the experimental results in \textbf{b}.
\textbf{d.} ARPES intensity maps along the c (x) axis measured near and below the excitonic transition temperature. Colored lines guide distinct dispersions of two valence bands (VB1 and VB2) and a conduction band (CB1). CB1 and VB1 open the excitonic band gap. 
\textbf{e.} dI/dV curve measured by STM and averaged for a unit cell.
\textbf{f.} Large-area STM topography images obtained at 300 and 100 mV, respectively.
Atomic scale defects appear at a lower bias as bright protrusions or points.
\textbf{g.} dI/dV maps obtained at -25 and 40 meV, respectively, within the band gap for the same area as \textbf{f}.
In-gap electronic states appear on bright defects of the topography. Scale bars in \textbf{f} and \textbf{g} and their inset represent 20 nm and 2 nm, respectively. 
 }\label{Fig 1}
\end{figure*}

\begin{figure*}[htp!]%
\centering
\includegraphics[width=0.9\textwidth]{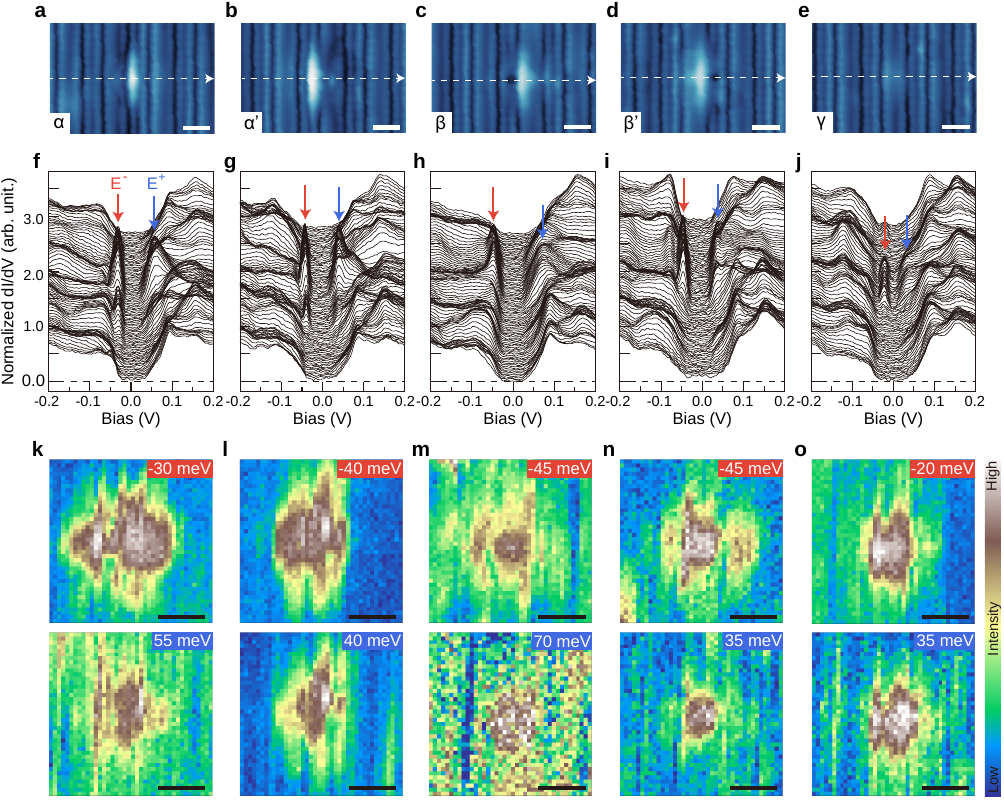}
\caption{\textbf{Identification of defects with in-gap bound states.}
\textbf{a-e.} STM topography images (sample bias of 100 mV) of defects exhibiting in-gap states with 1 nm scale bars.
Images for $\alpha$ and $\alpha$' ($\beta$ and $\beta$') are mirror symmetric.
\textbf{f-j.} dI/dV STS data measured along white dashed arrows in \textbf{a-e}. Each spectrum is shifted vertically for clarity. 
The in-gap states in the filled ($E^-$) and empty ($E^+$) states are indicated by red and blue arrows, respectively.
\textbf{k-o.} Grid dI/dV maps on each defect measured at its in-gap state energies $E^-$ (top) and $E^+$ (bottom).
Scale bars represent 2 nm and the in-gap states are distributed over approximately 5 nm.
.
}\label{Fig 2}
\end{figure*}

\begin{figure*}[htp!]%
\centering
\includegraphics[width=0.9\textwidth]{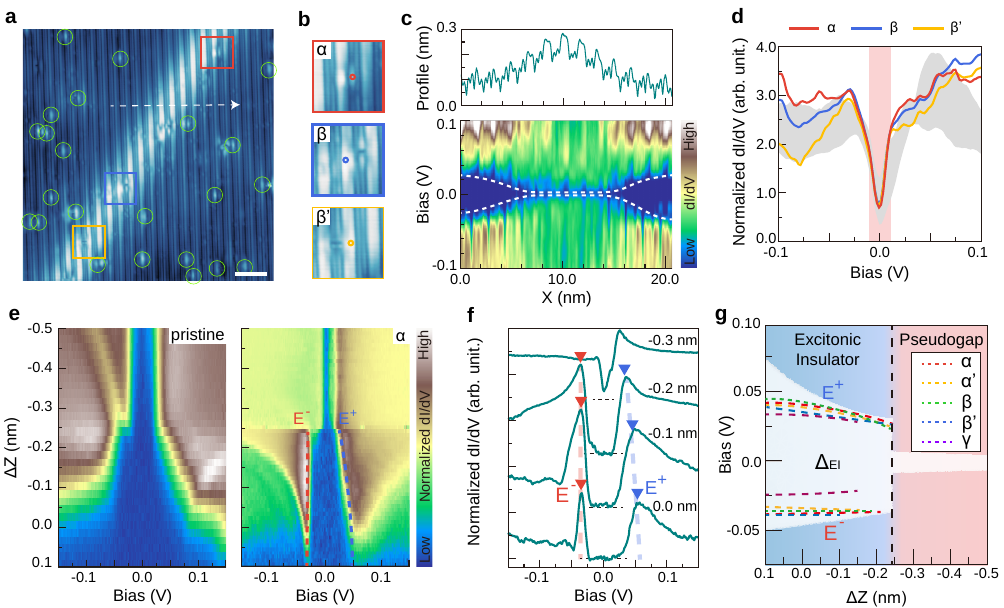}
\caption{
\textbf{a.} STM topography of a locally strained area (a stripe of bright protrusion). The scale bar represents 5 nm. 
Ta-vacancy defects are marked by circles or boxes. 
\textbf{b.} Enlarged topography of $\alpha$, $\beta$ and $\beta'$ defects within the strained area.
\textbf{c.} Height profile and dI/dV maps across the strained area (along the dashed arrows in \textbf{a}). The dI/dV map shows the excitonic gap closed.
\textbf{d.} Normalized dI/dV curves of the three defects in \textbf{b}. Similar dI/dV curves in the strained area away from defects distribute in the grey part. 
\textbf{e.} Changes of the dI/dV curves upon the variation of the tip height on (left) a pristine area and a defect $\alpha$ (right).
The zero tip height corresponds to the normal tunneling condition.
(Left) The turn-on of the dI/dV signal is indicated at the dashed lines, which indicate conservatively the gap edges and the gap edge spectral features are outside of the shaded region. (Right) The in-gap states are indicated by dashed lines. 
\textbf{f.} dI/dV curves at given ${\Delta}Z$ on defect $\alpha$ taken from \textbf{e}.
\textbf{g.} Summary of ${\Delta}Z$ sweeps on different defects. In-gap state energies are plotted against the gap edge positions of the pristine area (\textbf{e}). 
}\label{Fig 3}
\end{figure*}

\begin{figure*}[htp!]%
\centering
\includegraphics[width=0.9\textwidth]{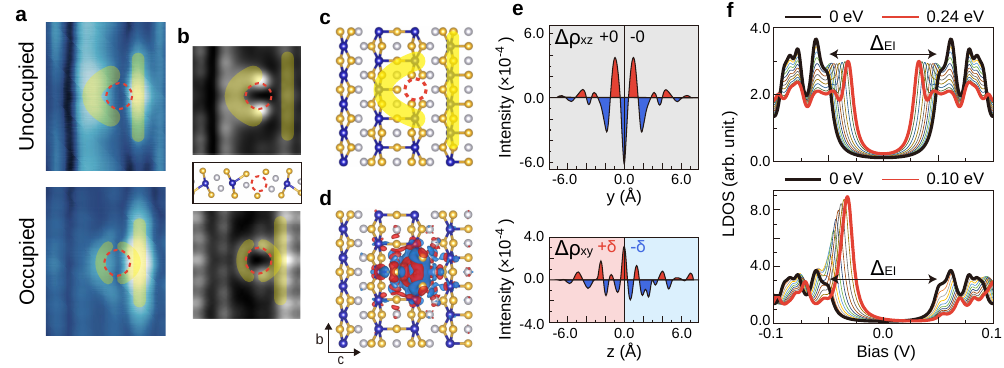}
\caption{
\textbf{a,b.} Enlarged STM topography and simulated STM images of a Ta vacancy  ($\alpha$ defect) at unoccupied (top) and occupied (bottom) states. The yellow shades indicate the major topographic features. 
The Ta vacancy position is shown in \textbf{a, b, c} by the red dashed lines.
\textbf{d.} Charge density difference due to the Ta vacancy from the pristine structure in the DFT calculation.
\textbf{e.} xz(xy)-plane averaged charge density difference in the vicinity of the Ta vacancy along the y (b) (top) and x-axis (bottom), indicating the charge dipole formation along the y-axis.
\textbf{f.} Numerical solution of 2D tight-binding Hamiltonian of the exciton insulator with the local charge dipole (top) and monopole (bottom) potential V, indicating the emergence of the symmetric twin in-gap state from the gap edges only in the case of charge dipole.
}\label{Fig 4}
\end{figure*}

\clearpage

\noindent{\Large \bf Supplementary Information for "Yu-Shiba-Rusinov bound states of exciton condensate"}

\noindent{SeongJin Kwon$^{1,2}$, Kyung-Hwan Jin$^{1,7}$, Jong Eun Han$^{3}$, Siwon Lee$^{1,2}$, ChoongJae Won$^{4,5}$, Sang-Wook Cheong$^{4,5,6}$, Han Woong Yeom$^{1,2\dagger}$}

\maketitle

\begin{affiliations}
 \item Center for Artificial Low Dimensional Electronic Systems, Institute for Basic Science, Pohang 37673, Korea.
 \item Department of Physics, Pohang University of Science and Technology, Pohang 37673, Korea.
 \item Department of Physics, University at Buffalo, State University of New York, Buffalo, New York 14260, USA.
 \item Laboratory for Pohang Emergent Materials, POSTECH, Pohang 37673, Korea
 \item MPPC-CPM, Max Planck POSTECH/Korea Research Initiative, Pohang 37673, Korea
 \item Keck Center for Quantum Magnetism and Department of Physics and Astronomy, Piscataway, New Jersey 08854, United States
 \item Department of Physics and Research Institute of Physics and Chemistry, Jeonbuk National University, Jeonju 54896, Republic of Korea
 
 $^{\dagger}$yeom@postech.ac.kr
 
\end{affiliations}

\clearpage

\null
\vfill
\begin{figure*}[htp!]%
\centering
\includegraphics[width=0.9\textwidth,clip]{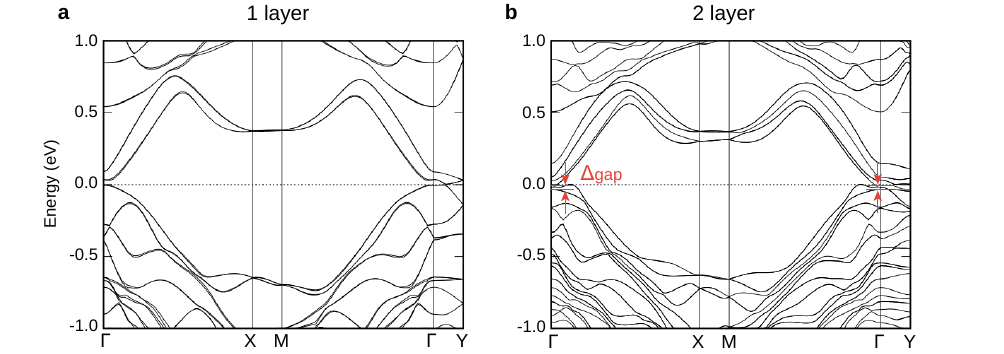}
\caption{\textbf{a,b.} DFT-calculated electronic band structure of monolayer (a) and bi-layer (b) Ta$_2$Pd$_3$Te$_5$. The band gap from the calculation is much smaller than the experimental excitonic gap, since electron-hole manybody interactions are not sufficiently included in DFT. These results are consistent with the previous DFT reports\cite{TPT1, TPT2, TPT3, TPT4, TPT9}}
\label{Fig S1}
\end{figure*} 
\vfill

\clearpage
\null
\vfill
\begin{figure*}[htp!]%
\centering
\includegraphics[width=0.9\textwidth,clip]{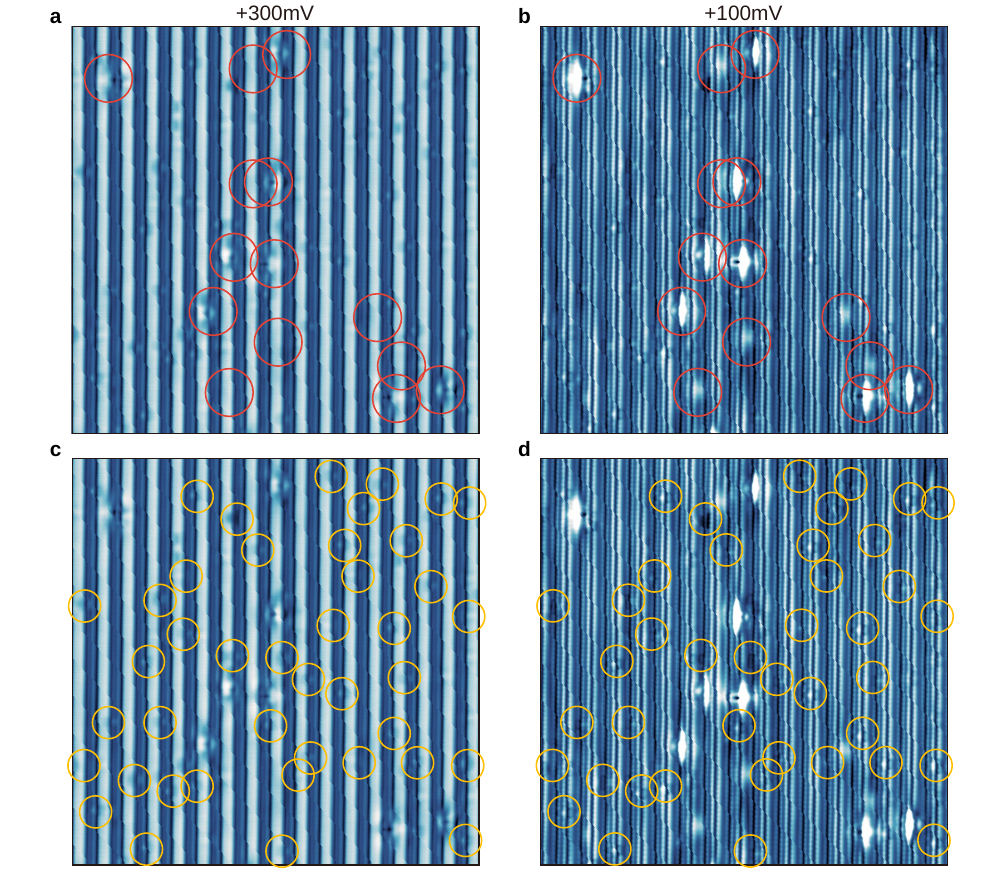}
\caption{\textbf{a-d.} STM topographies obtained at sample bias voltages +300 mV (left) and +100 mV (right), respectively, showing the distributions of non-trivial (red circle) and trivial (yellow) defects on a surface of Ta$_2$Pd$_3$Te$_5$ at 4.4 K. Top and bottom images are the same ones.
}
\label{Fig S2}
\end{figure*} 
\vfill

\clearpage
\null
\vfill
\begin{figure*}[htp!]%
\centering
\includegraphics[width=0.9\textwidth,clip]{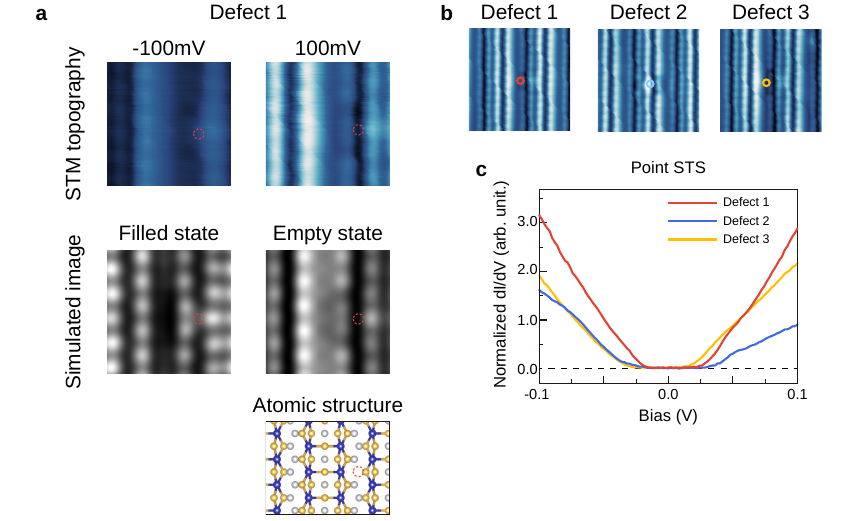}
\caption{\textbf{a.} STM topographies and DFT-simulated images of a Pd vacancy in the top layer of Ta$_2$Pd$_3$Te$_5$. The atomic structure of the vacancy (on the site indicated by the red circle) is given at the bottom, which is determined by DFT calculations.
\textbf{b.} Empty-state STM topographies of the different types of trivial defects and their STS (dI/dV) results obtained at colored points, respectively. All defects showed no in-gap states.}
\label{Fig S3}
\end{figure*} 
\vfill

\clearpage

\null
\vfill
\begin{figure*}[htp!]%
\centering
\includegraphics[width=0.9\textwidth,clip]{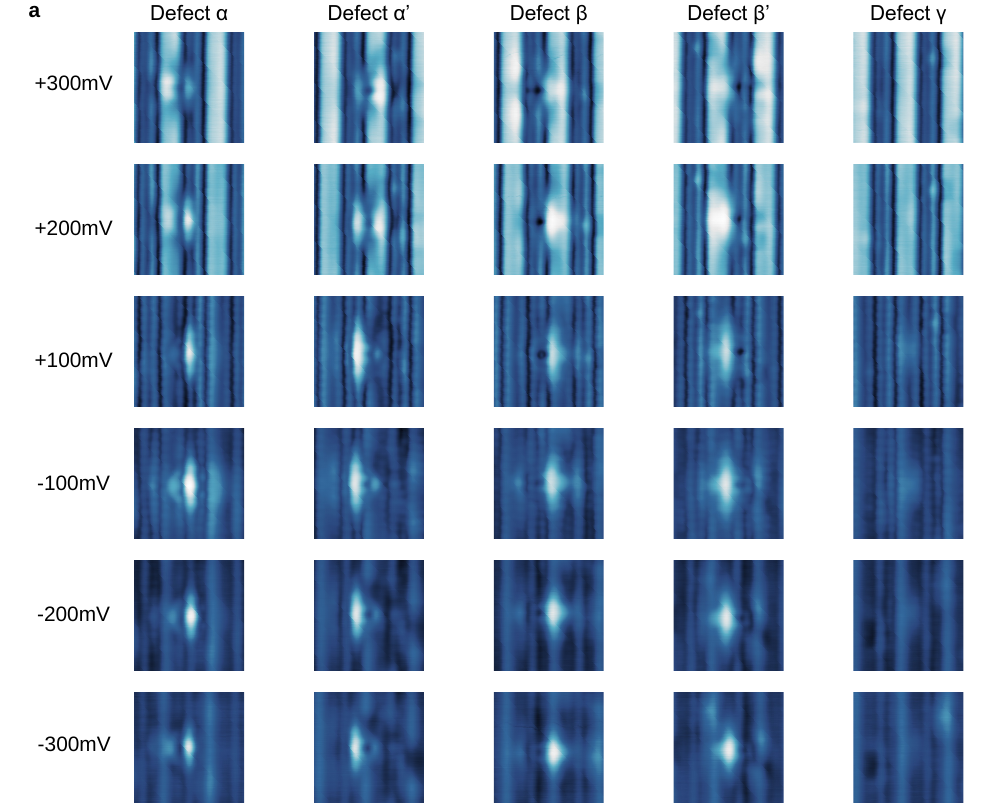}
\caption{\textbf{a.} Bias-dependent STM topographies of five different types of non-trivial Ta vacancy defects ($\alpha$, $\alpha'$, $\beta$, $\beta'$, and $\gamma$). The sample bias voltages are indicated, which range from -300 to +300 mV.}
\label{Fig S4}
\end{figure*} 
\vfill
\clearpage

\begin{figure*}[htp!]%
\centering
\includegraphics[width=0.9\textwidth,clip]{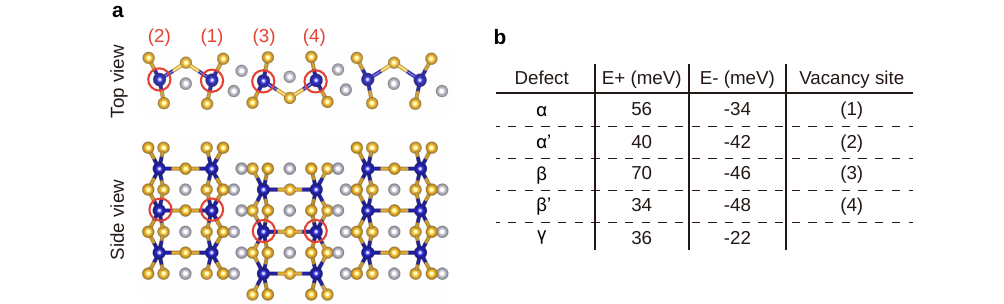}
\caption{In-gap state energies for different types of Ta vacancy defects. \textbf{a} Atomic structure of Ta$_2$Pd$_3$Te$_5$ (top and side view) with four distinct Ta vacancy sites (1), (2), (3), and (4). $\alpha$  and $\alpha'$ defects ($\beta$ and $\beta')$ occupy mirror-symmetric positions. \textbf{b} Measured in-gap state energies ($E^+$ and $E^-$) for each defect type. The $\gamma$ defect is thought to be due to Ta vacancies in the layer underneath.}
\label{Fig S5}
\end{figure*} 
\vfill

\null
\vfill
\begin{figure*}[htp!]%
\centering
\includegraphics[width=0.9\textwidth,clip]{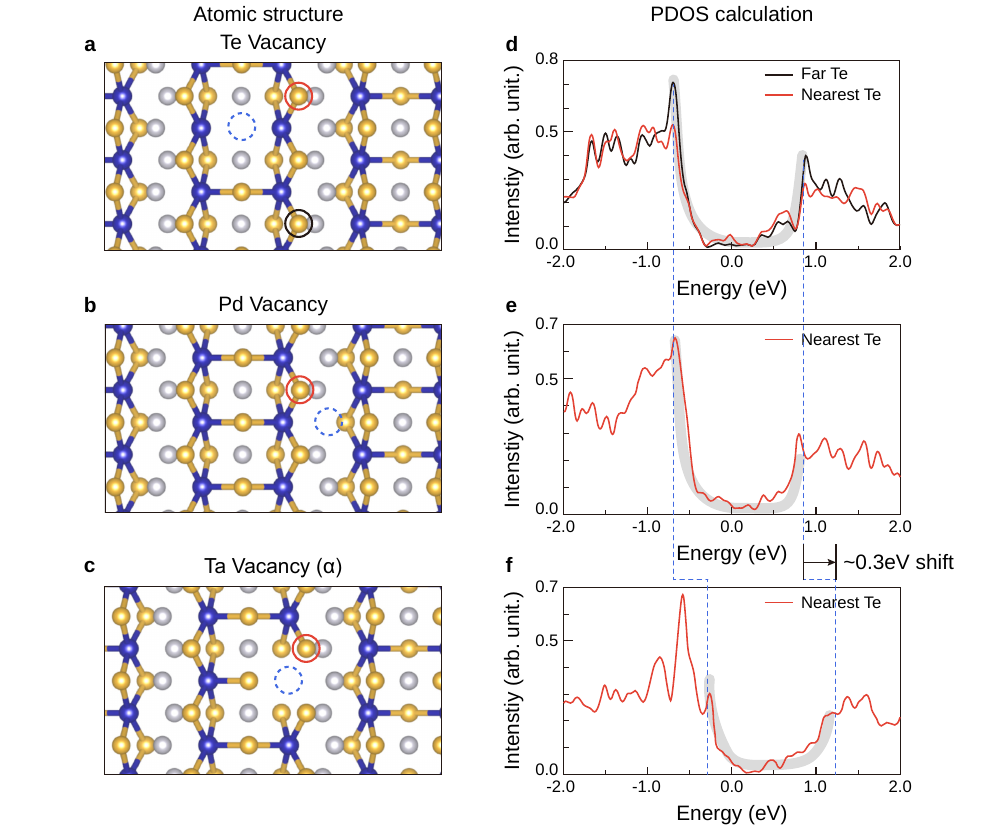}
\caption{\textbf{a,b,c.} DFT-calculated atomic structures of Te, Pd, and Ta vacancies (blue dotted circles). \textbf{d,e,f.} Corresponding Te partial density of states (PDOS) of a Te atom (red circles) near the defects, respectively. The Te PDOS of a pristine Te site is given for comparison in \textbf{d}. While Te and Pd vacancy do not show any significant shift of the DOS, the Ta vacancy causes a strong rigid shift of the whole DOS to +0.3 eV higher energy. This indicates that the Ta vacancy is locally charged (negatively on this particular Te atom site).}
\label{Fig S6}
\end{figure*} 
\vfill
\clearpage

\null
\vfill
\begin{figure*}[htp!]%
\centering
\includegraphics[width=0.9\textwidth,clip]{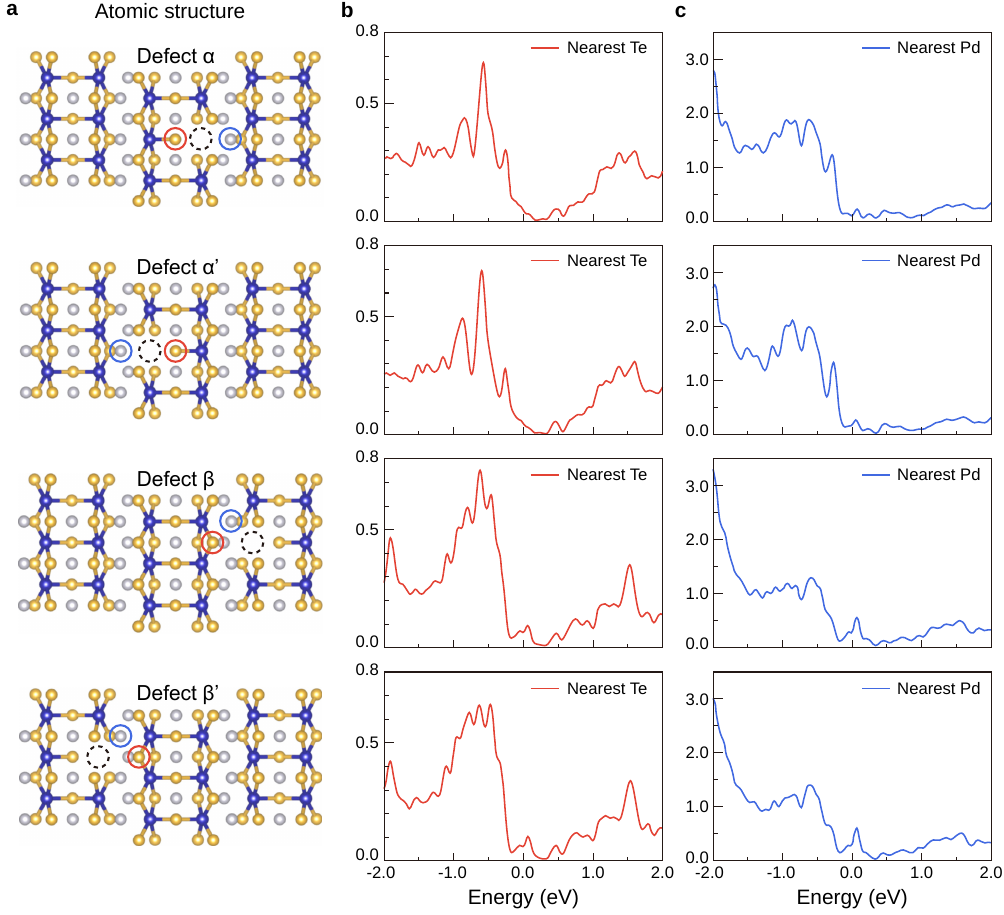}
\caption{\textbf{a.} DFT-calculated atomic structures of Ta vacancies (black dotted circles). \textbf{b,c.} Corresponding partial density of states (PDOS) of a Te atom (red circles) and Pd atom (blue circles) near the Ta defects, respectively.}
\label{Fig S7}
\end{figure*} 
\vfill
\clearpage

\begin{figure*}[htp!]%
\centering
\includegraphics[width=0.9\textwidth,clip]{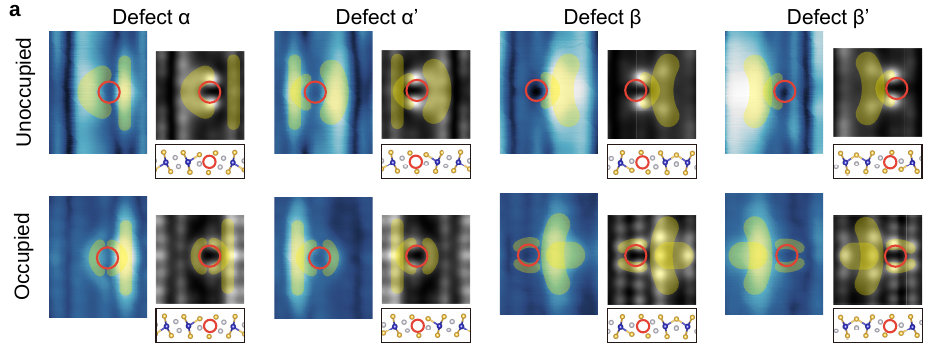}
\caption{\textbf{a.} Comparison between STM topographies and the corresponding DFT simulations (gray) with atomic structures at empty and filled states. The overall topographic contrast in the experiment agree reasonably with the simulated ones while some contrast as guided by red circles (Ta vacancy site) and yellow lines for the main features.}
\label{Fig S8}
\end{figure*} 
\vfill

\null
\vfill
\begin{figure*}[htp!]%
\centering
\includegraphics[width=0.9\textwidth,clip]{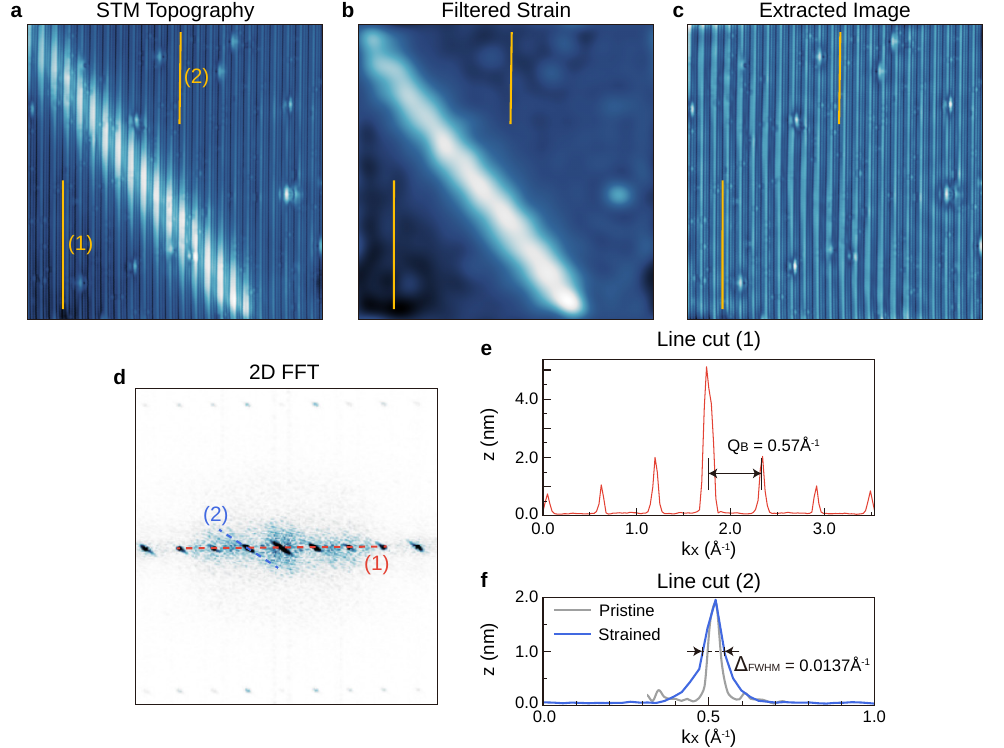}
\caption{\textbf{a,b,c.} STM Topography of strained area (same as in Fig. 4a), strain effect in the region extracted by 2D FFT filtering, and topography with the strain component removed. The angle difference 0.57 $^{\circ}$ between two yellow lines (1,2) indicates the local shearing of about 1 \%.  \textbf{d.} 2D Fast Fourier Transform of the STM topography in Fig. 4a, showing reciprocal space Bragg peaks clearly and their anisotropic broadening along the blue dashed lines, which reflects the local lattice strain with shear. 
\textbf{e.} Line profile (1) of the main Bragg peaks along the red dashed line. The Bragg vector of the lattice is determined as  $Q_B = 0.57\AA^{-1}$. \textbf{f.} Line profile (2) across a broadened Bragg peak along the blue dashed line, where the difference of full width at half maximum (FWHM) between strained area and pristine is measured to be 0.0137 $\AA^{-1}$, suggesting an average local strain variation of ${\Delta}_{FWHM}/Q_B \approx$ 2.4 \%.}
\label{Fig S9}
\end{figure*} 
\vfill

\clearpage



\null
\vfill
\begin{figure*}[htp!]%
\centering
\includegraphics[width=0.9\textwidth,clip]{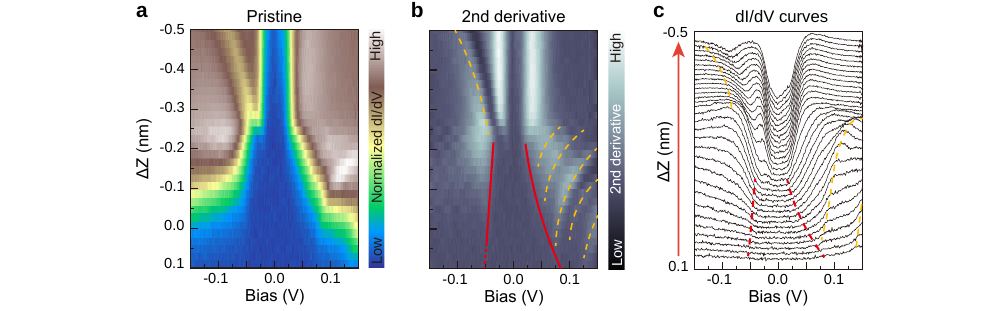}
\caption{Excitonic band gap tuning on a pristine surface area of Ta$_2$Pd$_3$Te$_5$ without defects with the tip-height control. \textbf{a} Color-scaled stack of dI/dV curves and \textbf{b} their second derivative curves ($d^2I/dV^2$) at different tip height (z). A few dI/dV curves at different tip height are also shown in \textbf{c}. Z=0 height represents the normal measurement height of the present STS experiments. The edges of valence and conduction bands LDOS's are marked by red solid or dashed lines. The panning out spectral features within the valence and conduction bands are attributed to the effect of strong electric field of the tip formed possibly through the Wannier-Stark ladder effect.}
\label{Fig S10}
\end{figure*} 
\vfill

\clearpage

\null
\vfill
\begin{figure*}[htp!]%
\centering
\includegraphics[width=0.9\textwidth,clip]{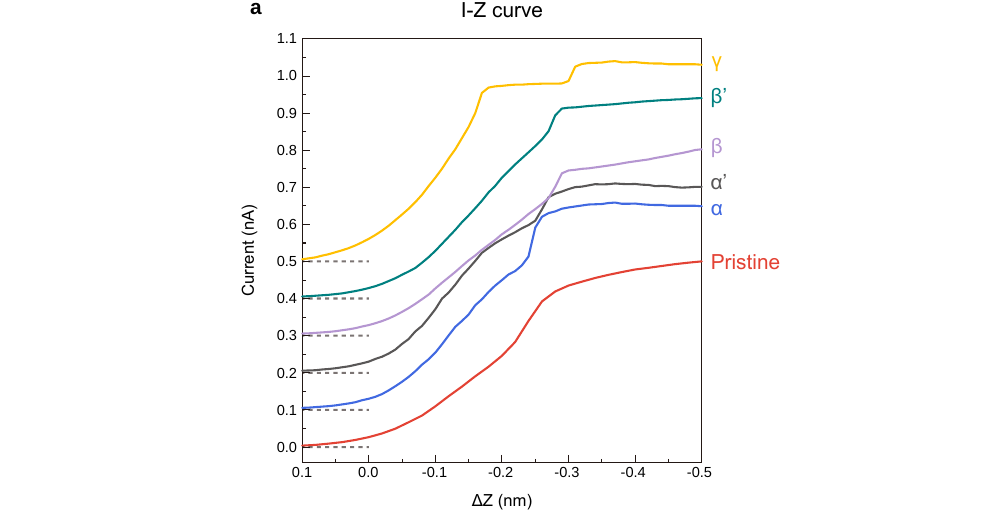}
\caption{\textbf{a.} Tunneling current (I) measured while the tip height is varied on a pristine surface area and on different Ta vacancy defects. The cureves are vertically shifted for visual clarity. At around the height of -0.3 nm the tunneling currents saturated to a consistent value of about 500 pA, which indicates the quantum contact regime.}
\label{Fig S11}
\end{figure*} 
\vfill

\null
\vfill
\begin{figure*}[htp!]%
\centering
\includegraphics[width=0.9\textwidth,clip]{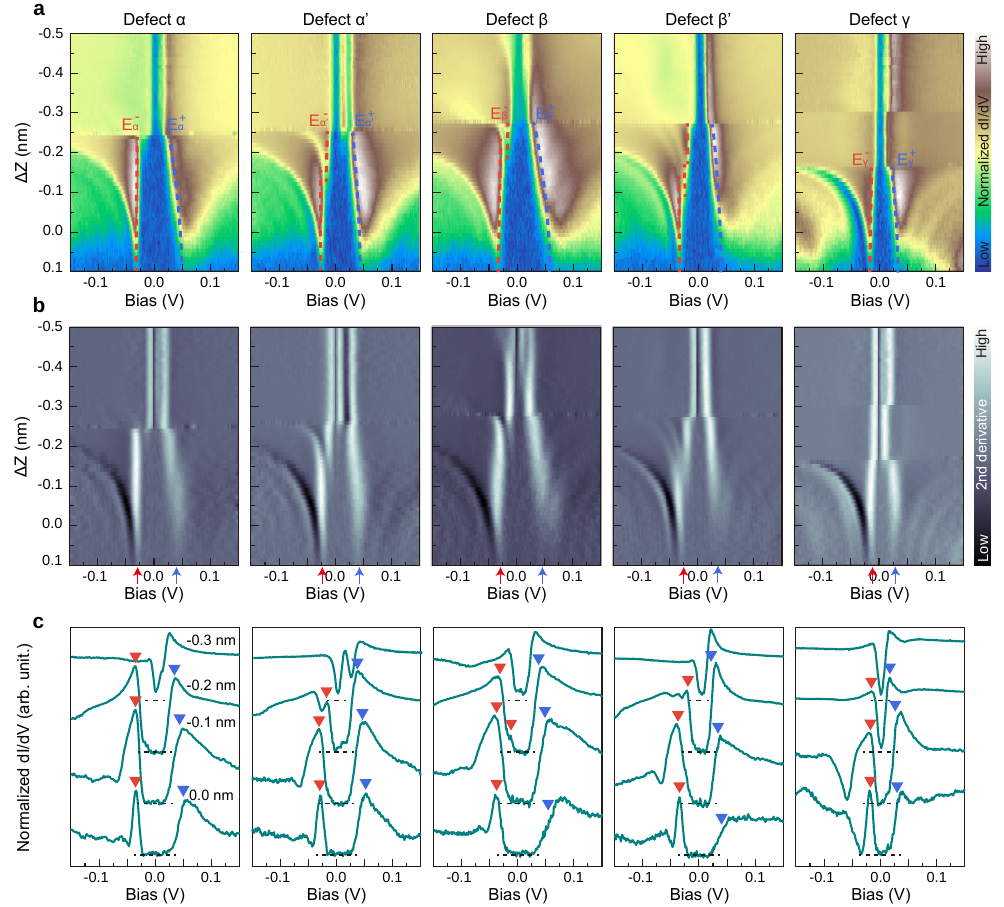}
\caption{Systematic gap tuning measurements on different types of Ta vacancy defects. \textbf{a} Normalized dI/dV data at different tip heights (${\Delta}z$) for each defect type. \textbf{b} Second-derivative of dI/dV ($d^2I/dV^2$) on each defects revealing the evolution of in-gap states as a function of the tip height. A few representative dI/dV cureves are given in \textbf{c}. All in-gap resonance peaks shift together with the excitonic gap edges, confirming their excitonic origin. The overall behaviors of the in-gap states and the excitonic energy gap are consistent among different defects. For the defect $\alpha$', an in-gap state appears within the pseudogap in the quantum contact regime, whose origin is not understood yet but can be related to the excitonic fluctuation (preformed excitons) represented by the pseudogap.}
\label{Fig S12}
\end{figure*} 
\vfill

\clearpage

\null
\vfill
\begin{figure*}[htp!]%
\centering
\includegraphics[width=0.9\textwidth,clip]{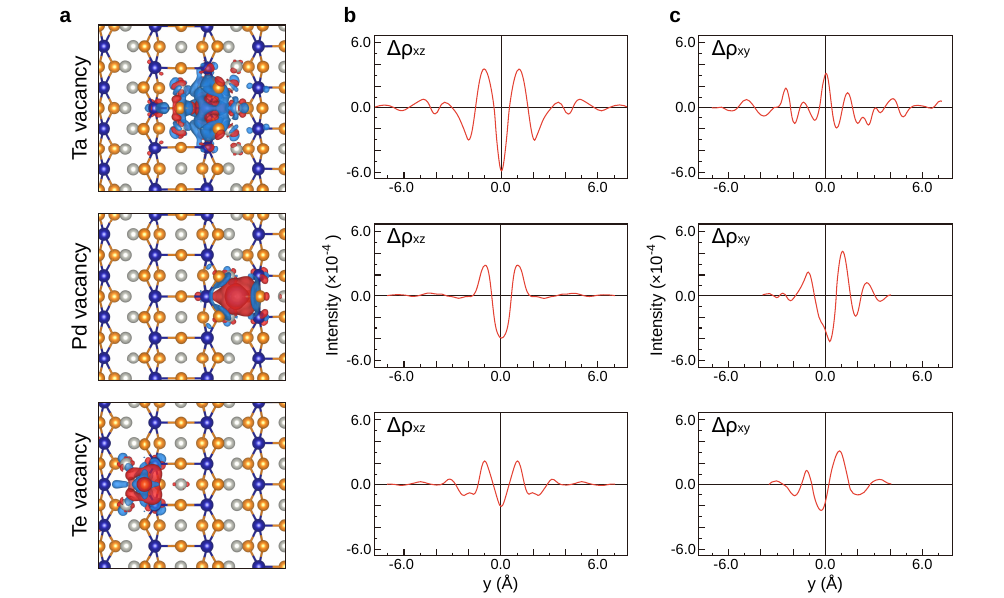}
\caption{\textbf{a.} Simulated charge polarization of Ta, Pd, and Te vacancy defects. The calculation shows $p_z=-0.45\times10^{-12}$ C${\cdot}$m dipole moment on Ta vacancy, while Pd and Te vacancy have $p_z=-0.02\times10^{-12}$ C${\cdot}$m and $p_z=-0.09\times10^{-12}$ C${\cdot}$m, respectively.}
\label{Fig S13}
\end{figure*} 
\vfill

\null
\vfill
\begin{figure*}[htp!]%
\centering
\includegraphics[width=0.9\textwidth,clip]{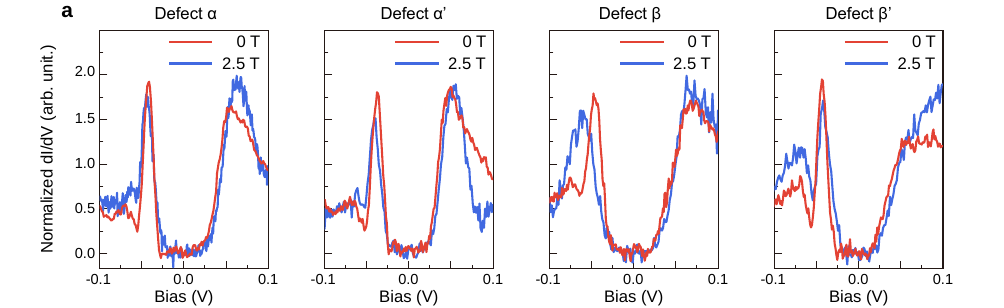}
\caption{\textbf{a.} Normalized dI/dV curves measured at magnetic field B = 0 T (red) and 2.5 T (blue) at a temperature of 1.26 K, respectively. The absence of Zeeman splitting up to 2.5 T for the in-gap states is not consistent with the behavior expected for a topological edge (defect) state.}
\label{Fig S14}
\end{figure*} 
\vfill

\clearpage

\null
\vfill
\begin{figure*}[htp!]%
\centering
\includegraphics[width=0.9\textwidth,clip]{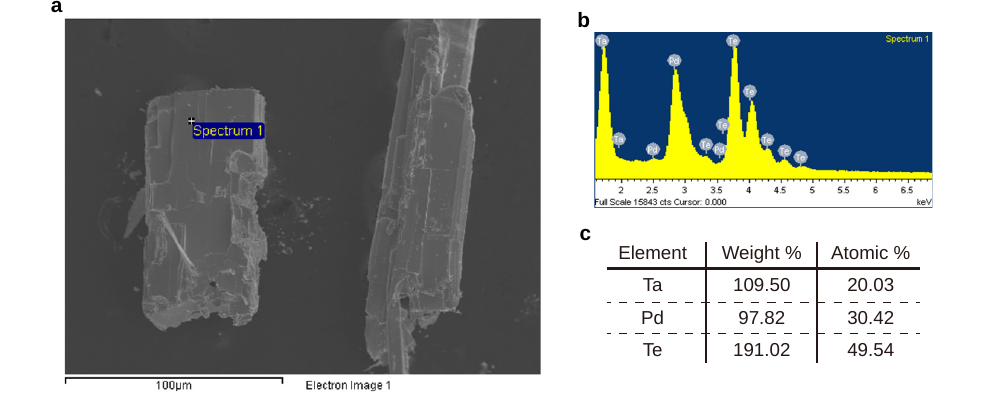}
\caption{\textbf{a} Electron microscopy image of a Ta$_2$Pd$_3$Te$_5$ crystal. 
\textbf{b} Energy dispersive X-ray spectroscopy analysis confirms the stoichiometry of Ta$_2$Pd$_3$Te$_5$ at a ratio of Ta : Pd : Te as 2 : 3 : 5 (\textbf{c}), respectively.}
\label{Fig S15}
\end{figure*} 

\begin{figure*}[htp!]%
\centering
\includegraphics[width=0.9\textwidth,clip]{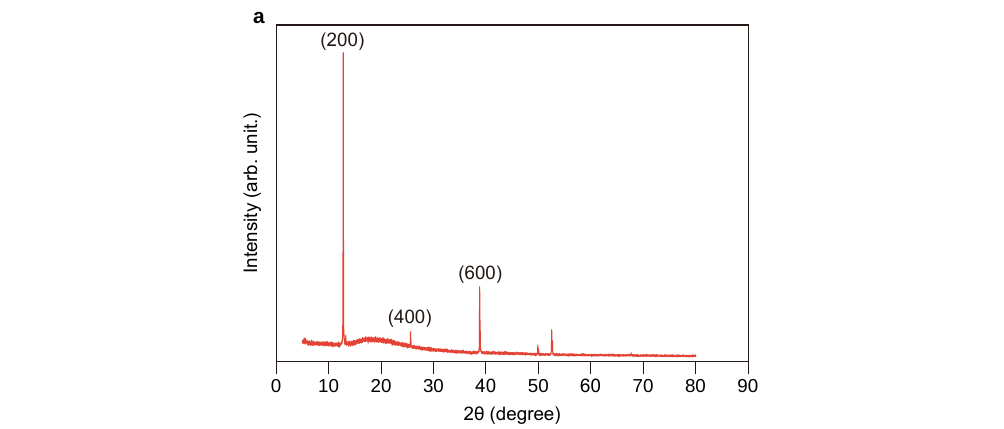}
\caption{\textbf{a} X-ray diffraction pattern of a Ta$_2$Pd$_3$Te$_5$ crystal, which is in good agreement with the previously reported data, confirming the integrity and precision of the synthesis.}
\label{Fig S16}
\end{figure*} 
\vfill

\clearpage




\end{document}